\documentclass[journal=nalefd,manuscript=article]{achemso}

\usepackage[version=3]{mhchem} 
\usepackage{xcolor}
\usepackage[normalem]{ulem}

\author{Rodrigo Becerra Silva}
\affiliation{Department of Physics and Astronomy, University of California, Davis, California 95616, United States}
\author{Xiang Yi}
\affiliation{School of Physics and Technology, Wuhan University, Wuhan 430072, China}
\author{Ziyi Song}
\affiliation{Department of Physics and Astronomy, University of California, Davis, California 95616, United States}
\author{Dong Yu}
\affiliation{Department of Physics and Astronomy, University of California, Davis, California 95616, United States}
\email{donyu@ucdavis.edu}

\title[exciton transport in TIs]
  {Field-Induced Dissociation Reveals Excitonic Long-Range Photocarrier Transport in Bulk-Insulating Bi$_2$Se$_3$ Nanoribbons}


\begin{document}

\begin{tocentry}
\includegraphics[scale=0.5]{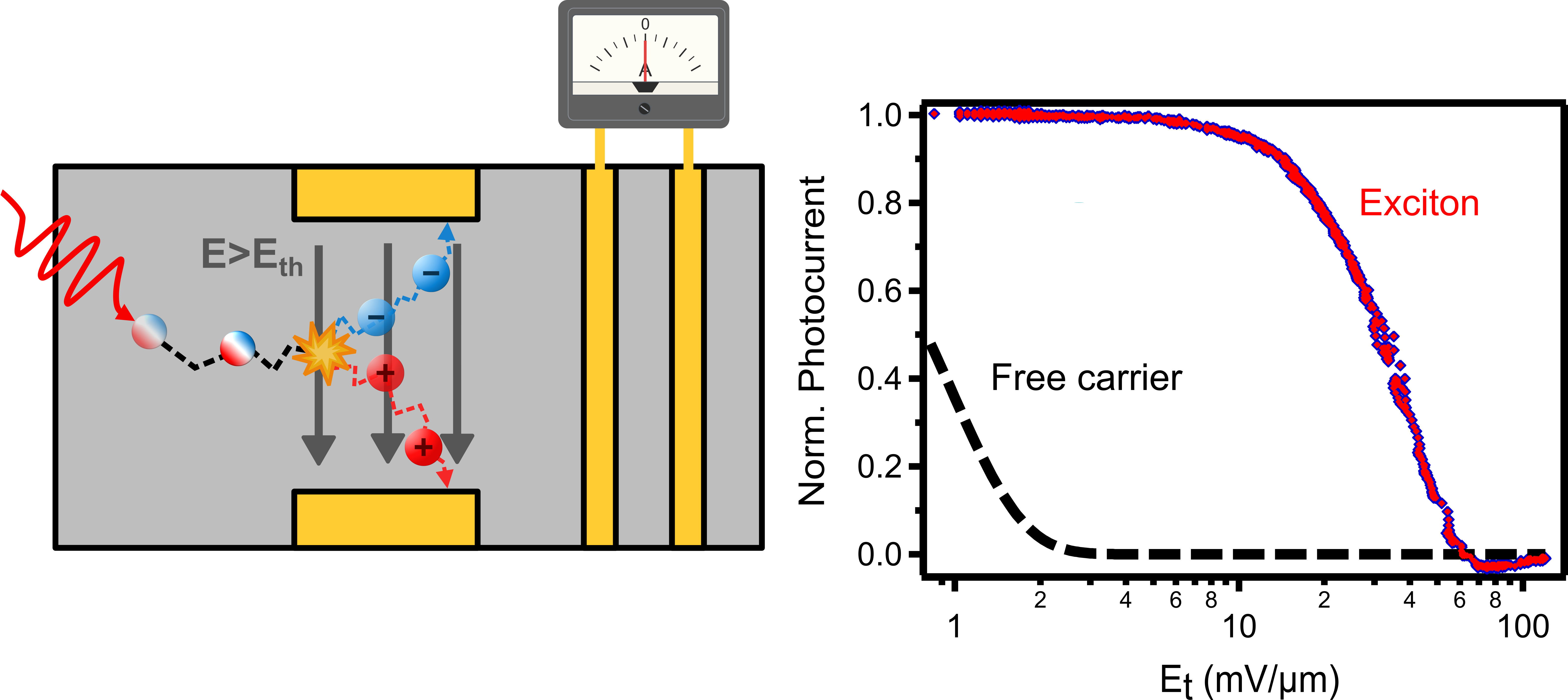}

\end{tocentry}


\begin{abstract}
Photoexcited charge carriers in topological insulators display anomalously long-range transport at cryogenic temperatures, but the underlying mechanism remains under debate. Here we use a transverse electric field as a discriminating probe of photocarrier dynamics in bulk-insulating Sb-doped Bi$_2$Se$_3$ nanoribbons, combining scanning photocurrent microscopy (SPCM) with ultrafast transient photovoltage (TPV) measurements. SPCM shows that suppression of the photocurrent requires a transverse electric field nearly 50 times larger than that needed to deflect free carriers, yet comparable to the expected exciton dissociation field. TPV measurements at 12 K and low excitation fluence reveal that the photocarrier diffusivity exceeds the value implied by the Einstein relation and the measured drift mobility by more than an order of magnitude, a bound that holds even at the lower end of the fitting uncertainty. Together, these observations indicate that the long-range photoresponse is not carried by free charge carriers but by a charge-neutral correlated state.

\end{abstract}


\newpage


Exciton condensation, the macroscopic occupation of a bound and coherent electron-hole state, has been pursued for more than half a century as a paradigmatic example of bosonic many-body order in solids.\cite{jerome1967} Early luminescence measurements in quantum wells provided the first signatures of excitonic condensation\cite{snoke2002long} and recent efforts have shifted toward bilayer systems made of two-dimensional (2D) materials such as graphene \cite{perali2013high, li2017excitonic, liu2017quantum} and transition metal dichalcogenides (TMDs).\cite{moon2025exciton} Topological insulators (TIs) have also emerged as a distinct and complementary platform: TIs have been proposed theoretically as hosts of topological exciton condensates,\cite{seradjeh2009,triola2017,pertsova2018} in which photoexcitation populates nonequilibrium excitons in the surface Dirac cone. On the experimental side, time- and angle-resolved photoemission has identified spin-polarized excitons in Bi$_2$Te$_3$.\cite{mori2023spin, mori2025possible} In bulk-insulating Sb-doped Bi$_2$Se$_3$ exhibiting nanosecond recombination lifetimes,\cite{gross2021nanosecond} we have recently reported photocarrier transport that extends to millimeter length scales\cite{hou2019millimetre} together with superdiffusive propagation.\cite{becerra2025superdiffusion} These behaviors are naturally accommodated within an exciton-condensate framework, but the supporting evidence to date remains indirect.

A conclusive realization of exciton condensation has remained elusive, in large part because the defining feature of an exciton, its charge neutrality, renders it invisible to the standard probes of condensed-matter physics. Conventional electrical transport couples to charged excitations and is, to leading order, blind to a neutral correlated state in the same sample.\cite{moon2025exciton} Establishing exciton condensation therefore requires either direct spectroscopic access to the condensate gap or transport signatures designed specifically to discriminate between charged and neutral carriers. Spectroscopic routes, however, are difficult to deploy across many of the leading platforms: bilayer systems host equilibrium interlayer excitons that carry weak or forbidden optical transitions, while in TIs such as Bi$_2$Se$_3$ the bulk band gap lies in the mid-infrared and luminescence is intrinsically weak. Transport-based approaches have therefore carried much of the experimental burden. Some of the most compelling experimental evidence to date comes from bilayer quantum Hall systems, where the absence of a Hall response under counterflow currents, a direct consequence of the vanishing Lorentz force on charge-neutral exciton pairs, has provided a definitive signature of exciton transport.\cite{nandi2012,finck2011} More recently, equilibrium excitonic insulator behavior has been reported in InAs/GaSb bilayers\cite{du2017, yu2018anomalously} and in TMD heterostructures hosting interlayer excitons.\cite{wang2019} A common thread across these systems is that the most direct evidence for condensation comes from observables sensitive to charge neutrality itself---Hall response, counterflow, and Coulomb drag.

\begin{figure}
    \centering
    \includegraphics[width=1\linewidth]{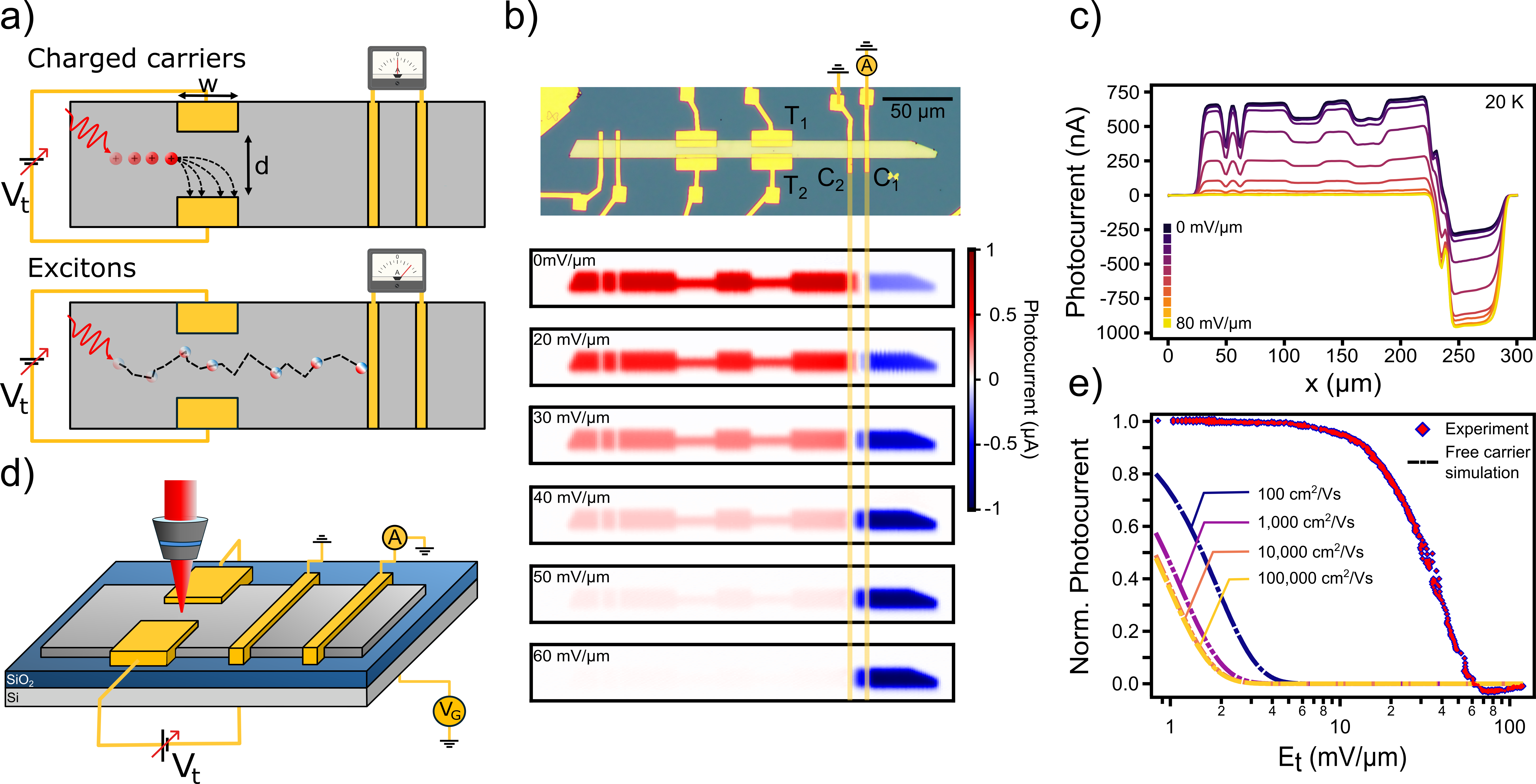}
    \caption{Transverse field dependent photocurrent in device D1 at 20 K. (a) Schematic of the device geometry with a pair of transverse contacts separated by a distance $d$. Charged photoexcited carriers are deflected by the transverse electric field, whereas excitons can transit through. (b) Optical image of the device together with SPCM maps acquired at different transverse electric-field strengths using a 5.3 $\mu$W laser at 690 nm. All maps are plotted using the same color scale. (c) Line cuts along the nanoribbon axis extracted from the center of the SPCM maps shown in (b), showing the evolution of the photocurrent profile with applied transverse field. (d) Schematic of the SPCM setup. (e) Experimentally measured, normalized photocurrent (PC) as a function of transverse field, plotted against the simulation results based on diffusion and drift of free carriers with several mobility settings. The free carrier model predicts a much lower field strength for photocurrent suppression, inconsistent with the experimental results. The photoexcitation is fixed at the center of the transverse field region in both experiment and simulation. }
    \label{fig:SPCM}
\end{figure}

Here we implement a direct test of charge neutrality on the long-range photoresponse of Sb-doped Bi$_2$Se$_3$ nanoribbons, using a transverse electric field to discriminate between charged and neutral photoexcited species. In the geometry we adopt, photoexcited carriers propagating along the nanoribbon are subjected to an electrostatic force perpendicular to their motion: free charged carriers are deflected and the photocurrent is suppressed, whereas a charge-neutral exciton experiences no net force on its center of mass and propagates undeflected [Figure \ref{fig:SPCM}(a)], up to a threshold field inducing excitonic dissociation. The threshold field at which the photocurrent collapses therefore reports directly on the charge of the photoexcited species, providing an analog of the absent-Hall-response argument that established neutral exciton transport in bilayer quantum Hall systems. Under transverse bias, we combine scanning photocurrent microscopy (SPCM) with ultrafast transient photovoltage (TPV) measurements, the latter providing an independent, in situ determination of the photocarrier diffusivity  with picosecond time resolution, for comparison with the drift mobility via the Einstein relation. Together, these complementary probes provide quantitative evidence that the long-range photoresponse in bulk-insulating Bi$_2$Se$_3$ is inconsistent with free charge carriers and is instead consistent with a charge-neutral correlated state.

Single crystalline Bi$_{2-x}$Sb$_x$Se$_3$ nanoribbons were grown by chemical vapor deposition (CVD) with $x$ = 0.3--0.4 determined by energy-dispersive X-ray spectroscopy (EDS) \cite{kong2010topological,hou2019millimetre}. Sb substitution compensates the residual bulk donors of Bi$_2$Se$_3$, suppressing bulk conduction and tuning the Fermi level into the bulk band gap. Single-nanoribbon field-effect transistors (FETs) with 100-200 nm thickness were fabricated by electron-beam lithography with 5 nm Cr / 295 nm Au contacts, and exhibited linear current-voltage characteristics with negligible contact resistance. We first present the SPCM results, which spatially map the photocurrent and provide detailed information on photocarrier transport.\cite{fu2011electrothermal,graham2013scanning, mcclintock2020temperature, wang2023spatially, kierdaszuk2026evidence} The SPCM setup is shown in Figure \ref{fig:SPCM}(d): a focused laser locally photoexcites the nanoribbon while the short-circuit photocurrent is monitored through the charge collection contacts (C1 and C2). Strikingly, at 20 K the photocurrent remains flat as the excitation spot is scanned along the 270 $\mu$m-long nanoribbon [Figure \ref{fig:SPCM}(b), (c)], while the internal quantum efficiency (IQE, defined as electrons collected per absorbed photon) reaches 30\%, similar to our previous report.\cite{hou2019millimetre} Note that the dips in photocurrent are caused by the partial light blocking by the contacts [Figure \ref{fig:SPCM}(c)]. The non-local efficient transport persists up to 60 K and the photocurrent then decays faster above 100 K (Figure S2 in Supporting Information).

We then apply an external electric field transverse to the carrier propagation direction using a pair of side contacts (T1 and T2) separated by a gap of $d$ = 5 $\mu$m and extending over a length of $w$ = 30 $\mu$m along the channel [Figure \ref{fig:SPCM}(b)]. The longitudinal current-collection contacts (C1 and C2) are held at ground, while the two transverse contacts are biased to a potential difference $V_\mathrm{t}$ by a floating DC supply. Because the collection contacts are grounded and symmetrically placed, the floating supply self-references so that the two transverse contacts sit at approximately +$V_\mathrm{t}$/2 and -$V_\mathrm{t}$/2 relative to ground, as confirmed experimentally. The drive current then flows predominantly between the transverse contacts. 

The metal--TI interface at each side contact presents a region of strong band bending and a high density of states, where photocarriers recombine far more rapidly than in the bulk of the nanoribbon and are removed from the longitudinal collection circuit. Even at zero transverse bias, carriers that diffuse to the side contacts are lost in this way, producing only a modest reduction of the collected photocurrent relative to a device without side contacts. A transverse field biases the otherwise isotropic transverse motion of free carriers: once the field-driven drift toward the nearest side contact, a distance $d/2$ away, outpaces thermal diffusion, the carriers are swept to the recombination sink at a much faster rate, and the longitudinal photocurrent is fully suppressed.

To establish a quantitative benchmark for the response of free photocarriers, we first estimate the threshold transverse electric field  $E_{\mathrm{th}}$ required to deflect them out of the conducting channel. Photocarriers diffuse across the half-gap $d/2$ in a characteristic diffusion time $t_\mathrm{diff} = (d/2)^2/D$, where $D$ is the diffusivity. Under a transverse field $E_\mathrm{t}$, the carriers drift across the same distance in a time $t_\mathrm{drift} = (d/2)/(\mu E_\mathrm{t})$, where $\mu$ is the carrier mobility. The field becomes important when the drift time is comparable to the diffusion time, $t_\mathrm{drift} = t_\mathrm{diff}$, which yields a threshold field,
\begin{equation}
  E_{\mathrm{th}} = \frac{2D}{\mu d}  = \frac{2 k_\mathrm{B} T}{qd} 
  \label{eq:Eth}
\end{equation}
where $k_\mathrm{B}$ is the Boltzmann constant and $q$ is the electron charge; the last equality invokes the Einstein relation $D = \mu k_\mathrm{B} T / q$. The diffusivity and mobility cancel: $E_{\mathrm{th}}$ is set entirely by the device geometry and the temperature, with no dependence on the carrier mobility. At $T$ = 20 K, Eq.~(\ref{eq:Eth}) gives $E_{\mathrm{th}} \approx$ 0.7 mV/$\mu$m. A more rigorous numerical simulation by solving the 2D drift-diffusion equations for free carriers confirms this simple estimate and gives a similar threshold field [dot-dashed curves in Figure \ref{fig:SPCM}(e), see details in SI]. The several-mobility curves demonstrate the insensitivity of the simulated threshold to mobility. 

In sharp contrast, the measured photocurrent remains essentially unchanged as the transverse field is increased to $E_\mathrm{t} \approx$ 10 mV/$\mu$m. Beyond this onset, the photocurrent drops steeply, reaching zero by $E_\mathrm{t} \approx$ 60 mV/$\mu$m, and remains fully suppressed at higher $E_\mathrm{t}$. The threshold and the suppression are independent of bias polarity (Figure S4 in SI). Taking the 50\% crossing of the photocurrent as the threshold, we estimate $E_{\mathrm{th}} \approx$ 33 mV/$\mu$m, which is about 47 times the free-carrier benchmark of Eq.~(\ref{eq:Eth}). We ruled out the potential artifacts including bias-induced Joule heating and slight misalignment of transverse contacts (see SI). Among the five devices we measured, $E_{\mathrm{th}}$ ranges from 11 to 33 mV/$\mu$m (Table S1 in SI). The variation is likely caused by the difference in Sb doping and defect distributions among the nanoribbons that can affect their chemical potential.~\cite{hou2020nonlocal}

A discrepancy of this magnitude cannot be reconciled with any free-charge-carrier picture: the cancellation of $\mu$ and $D$ in Eq.~(\ref{eq:Eth}) makes $E_{\mathrm{th}}$ independent of mobility, leaving no parameter that could account for the deviation. The observation is, however, naturally accommodated if the long-range photoresponse is carried by charge-neutral excitons. A uniform transverse field exerts no net force on the exciton center of mass, so the photoresponse is suppressed only once the field is strong enough to dissociate the exciton. The exciton binding energy and Bohr radius in Sb-doped Bi$_2$Se$_3$ are unknown but a rough theoretical estimate gives $E_\mathrm{b}$ = 3 meV and $a_\mathrm{B}$ = 110 nm.\cite{hou2019millimetre}  Equating the field-induced potential drop across an exciton to its binding energy yields an ionization field $E_\mathrm{diss} \sim E_\mathrm{b}/(q \ a_\mathrm{B}) \approx$ 27 mV/$\mu$m, consistent with the observed onset of suppression.

The spatial structure of the photocurrent suppression, resolved by SPCM, further supports the exciton interpretation. Suppression occurs not only when the laser is positioned beyond the transverse-field region, so that the photoexcited species must traverse the biased section to reach the collection contacts, but also when the laser sits between the field region and the collection contacts, where no such traversal is required. The suppression is in fact uniform for any laser position on the field-region side of the collection contacts [Figure \ref{fig:SPCM}(b), (c)]. Within the exciton picture, the biased region acts as a localized dissociator: excitons diffusing in from either direction are field-ionized upon entry, and the resulting free carriers are deflected out of the channel before collection. The uniformity in laser position requires that excitons reach the dissociation region from anywhere in the channel, consistent with the millimeter-scale propagation established previously.\cite{hou2019millimetre}

A second observation is more striking. When the laser is positioned on the opposite side of the collection contacts, away from the transverse-field region [$x > 240 \ \mu$m in Figure \ref{fig:SPCM}(c)], the photocurrent is enhanced rather than suppressed under transverse bias. The two effects are quantitatively linked: at the bias where the field-region-side photocurrent is fully suppressed, the opposite-side photocurrent is enhanced by an equal amount, corresponding to a $\sim$100\% increase relative to its zero-bias value. The enhancement, like the suppression, is independent of bias polarity. Together, these observations imply a global, bias-polarity-symmetric shift of the photocurrent under transverse bias, rather than a local modulation tied to either contact. Such a global shift has no counterpart in a single-pass transport picture and instead requires that propagating excitons encounter the field region repeatedly over their lifetime: a multiple-pass scenario.

To understand the effect of the transverse bias on the photocurrent distributions determined by SPCM, we develop a one-dimensional steady-state diffusion model for exciton transport along the nanoribbon. The exciton density $n(x)$ satisfies the continuity equation
\begin{equation}
  \frac{d^2 n}{dx^2} - \frac{n}{L_\mathrm{d}^2}
  - s_\mathrm{T}\, n(x)\, \delta(x - x_\mathrm{T})
  = -\frac{1}{D}\,\delta(x - x_0),
  \label{eq:diffusion}
\end{equation}
where $L_\mathrm{d} = \sqrt{D\tau}$ is the exciton diffusion length, $\tau$ is the recombination lifetime, and $x_0$ is the laser excitation position. The only sink in the transport equation is the transverse contact pair T1/T2, modeled as localized loss of strength $s_\mathrm{T}$ (units of $\mu$m$^{-1}$) at the position $x_\mathrm{T}$ that increases with $|E_\mathrm{t}|$, representing the field-assisted exciton dissociation and the consequent free carriers sweeping out. Reflecting boundary conditions ($dn/dx = 0$) are imposed at both ends of the nanoribbon. 

The longitudinal collection contacts C1 and C2 are not treated as exciton sinks. Band bending at these contacts can dissociate some excitons, and a small fraction of the resulting carriers may recombine at the metal interface; our model assumes this capture rate is low and neglects it. The dominant role of C1 and C2 is instead to act as potential probes. Excitons in local equilibrium with a dilute free-carrier population shift the electrochemical potential of the channel in proportion to the local exciton density, and the current measured by the preamplifier is the resulting potential difference between the two probes divided by the channel resistance $R_\mathrm{ch}$ between them:
\begin{equation}
  I_{\mathrm{ph}} = \frac{\mu_\mathrm{ec}(x_\mathrm{C2}) - \mu_\mathrm{ec}(x_\mathrm{C1})}{q\,R_\mathrm{ch}} \propto \gamma_2\, n(x_\mathrm{C2}) - \gamma_1\, n(x_\mathrm{C1}),
  \label{eq:photocurrent}
\end{equation}
where $\gamma_{1,2}$ are the conversion coefficients relating exciton density to probe potential at the two contacts. The overall proportionality constant is left undetermined, as the microscopic conversion from exciton density to measured current is not known; only the ratio $\eta \equiv \gamma_1/\gamma_2$ affects the shape of the profiles. We note, however, that because the collection contacts sense rather than absorb excitons, the measured current is governed by the exciton-density difference across the pair rather than by a small net capture rate; a high apparent IQE is therefore compatible with the long diffusion length and low bulk recombination in this regime.

We solve Eq.~(\ref{eq:diffusion}) numerically on a finite-difference grid with an exciton diffusion length $L_\mathrm{d} = 900$~$\mu$m, consistent with the value reported in Ref.~\citenum{hou2019millimetre}. A single asymmetry parameter $\eta = 0.997$, reflecting a slight ($\sim$0.3\%) difference in conversion efficiency between the two contacts, reproduces the measured ratio of the photocurrent magnitudes on the two sides of the collection contacts at zero field. The simulated profiles depend only weakly on the assumed diffusion length: reducing $L_d$ from 900 to 500~$\mu$m leaves both the profile shape and the position independence of the field-induced shift unchanged (Figure~S7). Diffusion lengths below $\sim$100~$\mu$m, however, yield profiles localized near the collection contacts, incompatible with the observed plateau.

\begin{figure}
    \centering
    \includegraphics[width=1\linewidth]{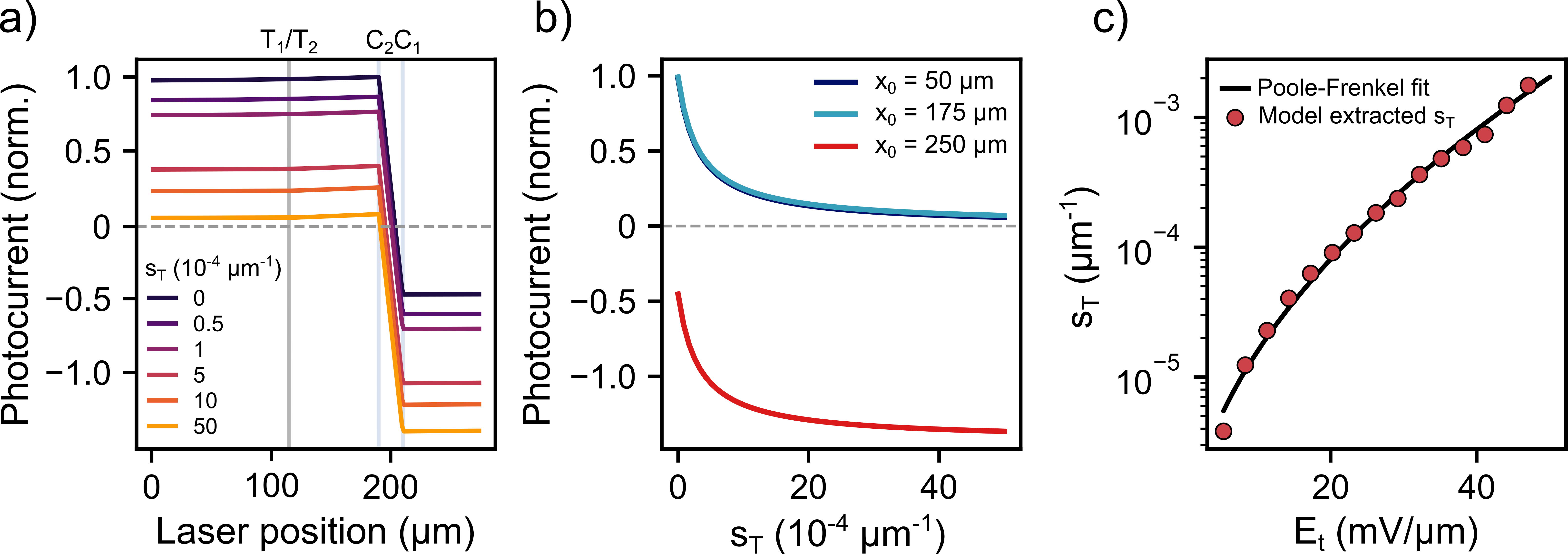}
    \caption{Simulation results of the multiple-pass exciton diffusion model. (a) Simulated photocurrent profiles as a function of laser position for increasing transverse sink strength $s_\mathrm{T}$. (b) Photocurrent as a function of $s_\mathrm{T}$ at representative laser positions. (c) Sink strength $s_\mathrm{T}$ extracted from the measured field-dependent photocurrent suppression, plotted against the transverse field $E_\mathrm{t}$. The solid line is a Poole-Frenkel fit using Eq. (\ref{eq:PF}).}
    \label{fig:model}
\end{figure}

Figure~\ref{fig:model}(a) shows the simulated photocurrent profiles as a function of laser position for increasing $s_\mathrm{T}$. At $s_\mathrm{T} = 0$ (no transverse bias), the photocurrent is positive to the left of C2 and negative to the right of C1, with the sign set by the exciton-density gradient across the contact pair. The photocurrent is nearly uniform within each region, consistent with the regime $L_\mathrm{d} \gg L$ where the exciton density is approximately position-independent. As $s_\mathrm{T}$ increases, the entire photocurrent profile shifts downward by a nearly position-independent amount, reproducing the key experimental observation. The model also captures the photocurrent vs. $s_\mathrm{T}$ at three representative laser positions [Figure~\ref{fig:model}(b)]: regardless of whether the excitation is to the left of T1/T2, between T1/T2 and C2, or to the right of C1, the photocurrent decreases at similar rates. The same mechanism accounts for the enhanced photocurrent magnitude when the laser is positioned to the right of C1: draining excitons at T1/T2 lowers the density more at the nearer probe C2 than at C1, steepening the gradient across the pair and driving the negative photocurrent further negative. The suppression on one side and the enhancement on the other are thus two faces of a single, polarity-symmetric global shift.

The transverse sink strength $s_\mathrm{T}$ increases with $|E_\mathrm{t}|$ because the transverse field promotes exciton dissociation at the T1/T2 contacts. This is a field-assisted (Poole-Frenkel-type) process in which the transverse field tilts the Coulomb potential binding the electron and hole and lowers the dissociation barrier,\cite{frenkel1938, liraz2022} so that
\begin{equation}
  s_\mathrm{T} \propto \exp\!\left(-\frac{E_\mathrm{b} - \beta\sqrt{E_\mathrm{t}}}{k_\mathrm{B}T}\right),
  \label{eq:PF}
\end{equation}
where $\beta$ is the Poole-Frenkel coefficient. Inverting the measured suppression through the model gives $s_\mathrm{T}(E_\mathrm{t})$ [Fig.~\ref{fig:model}(c)], which spans about three orders of magnitude and follows Eq.~(\ref{eq:PF}), confirming field-assisted dissociation of a Coulomb-bound electron-hole pair rather than a single-particle transport effect.


\begin{figure}
    \centering
    \includegraphics[width=1\linewidth]{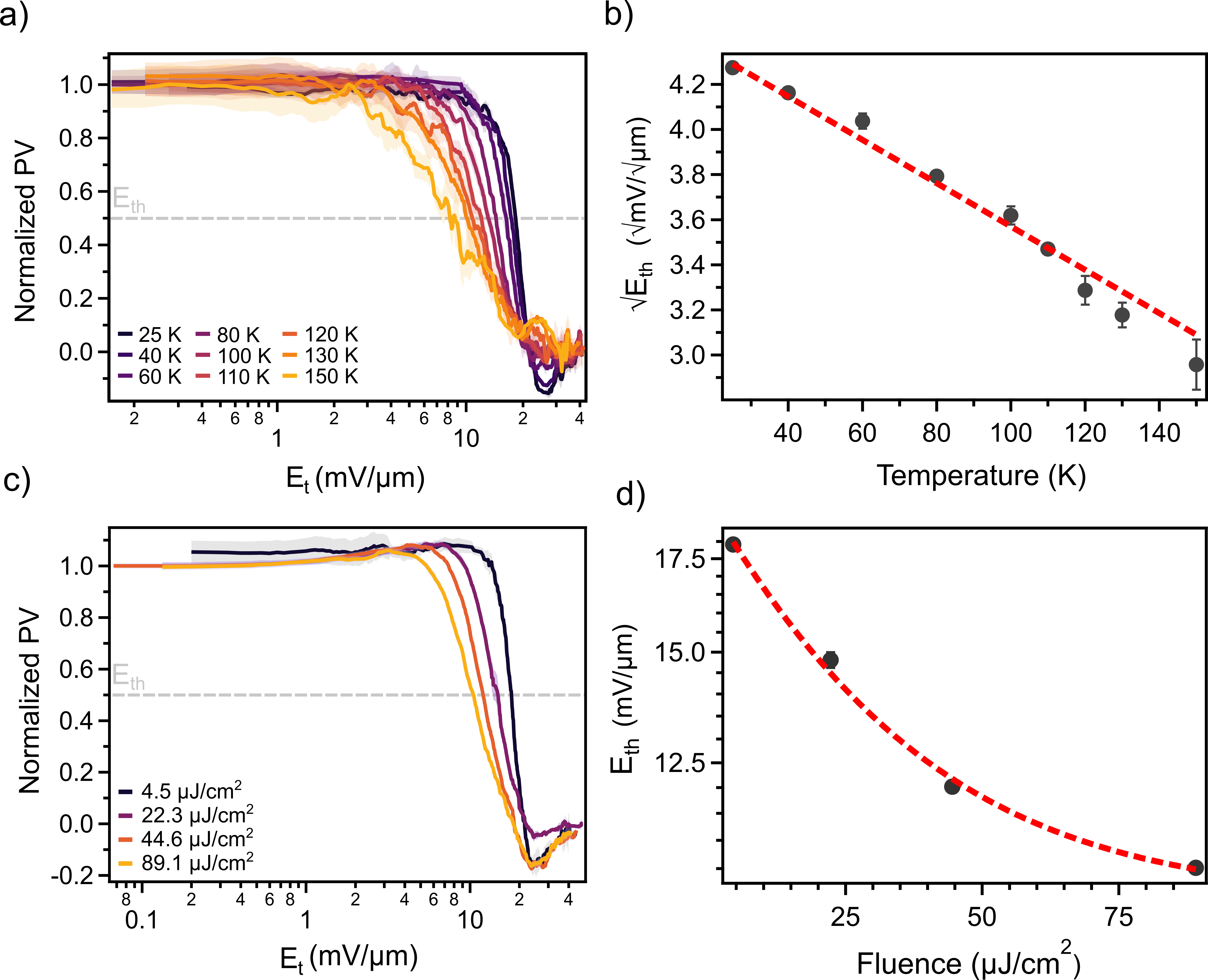}
    \caption{Temperature- and fluence-dependent threshold field in device D2. (a) Normalized photovoltage (PV) versus transverse field $E_\mathrm{t}$ for temperatures from 25 to 150 K, with the laser parked at the center of the transverse channel at a fluence of 4.8 $\mathrm{\mu J/cm^{2}}$. The threshold field $E_{\mathrm{th}}$ is defined as the 50\% crossing of the normalized PV (gray dashed line). Shaded bands denote the measurement uncertainty. (b) $E_{\mathrm{th}}$ extracted from (a) versus temperature, showing that $\sqrt(E_{\mathrm{th}})$ decreases linearly with temperature. (c) Same measurement as in (a), at 45 K, for laser fluences of 4.5-89.1 $\mathrm{\mu J/cm^{2}}$. (d) $E_{\mathrm{th}}$ extracted from (c) versus fluence, decreasing with increasing fluence (red dashed line: guide to the eye).}
    \label{fig:Eth-dep}
\end{figure}

We next examine how the threshold field evolves with temperature and excitation fluence (Figure \ref{fig:Eth-dep}). Parking the laser at the center of the transverse channel and sweeping the transverse bias reproduces the same field-independent plateau followed by a sharp collapse seen in the spatial scans, allowing $E_{\mathrm{th}}$ to be extracted as the 50\% crossing of the normalized photovoltage at each temperature [Figure \ref{fig:Eth-dep}(a)]. At threshold the dissociation rate of Eq.~(\ref{eq:PF}) is fixed, giving $\sqrt{E_{\mathrm{th}}} = \frac{E_\mathrm{b}}{\beta} - \frac{C\,k_\mathrm{B}}{\beta}\,T$, where $C$ is a constant. Figure~\ref{fig:Eth-dep}(b) confirms this predicted linear dependence of $\sqrt{E_{\mathrm{th}}}$ on $T$ from 25 to 150~K, an independent temperature-domain signature of the same
field-assisted dissociation: added thermal energy helps unbind the pair, so a weaker field suffices as $T$ rises. In contrast, the free-carrier benchmark of Eq.~(\ref{eq:Eth}) increases with $T$, opposite to the measured trend.

A closely related trend appears with excitation fluence. At fixed temperature (45 K), increasing the fluence from 4.5 to 89.1 $\mathrm{\mu J/cm^{2}}$ shifts the collapse to progressively lower fields, and $E_{\mathrm{th}}$ decreases correspondingly [Figure \ref{fig:Eth-dep}(c),(d)]. This is consistent with the same physical picture: at higher fluence the larger photoexcited carrier density weakens the electron-hole binding, lowering the field needed to complete the dissociation. Unlike the temperature dependence, which follows the Poole-Frenkel form, the fluence dependence has no comparably simple functional form: it reflects the density dependence of the exciton binding energy and screening, whose quantitative extraction is beyond the scope of the present work.

\begin{figure}
     \centering
     \includegraphics[width=1\linewidth]{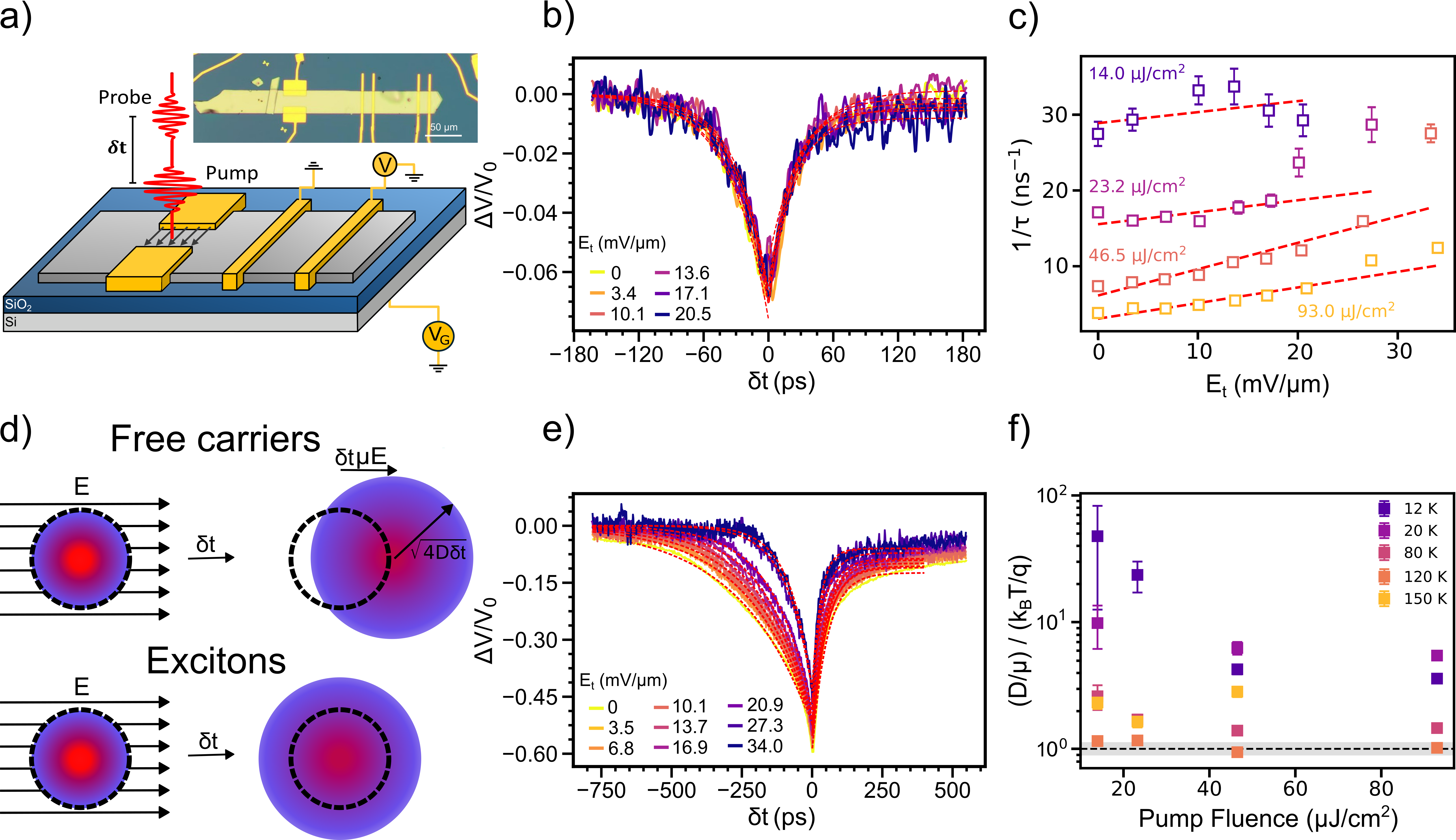}
     \caption{ Ultrafast transient photovoltage measurements under transverse field in device D2. (a) TPV setup, where a pulsed laser excites the device while a transverse electric field is applied across the channel. Inset: microscopic image of the device. (b,e) Normalized modulation of photovoltage versus delay measured at 14 and 93 $\mathrm{\mu J/cm^2}$ respectively, under various field strengths at 12 K. Positive (negative) delay means pump arrives later (earlier) than probe. (c) The inverse of recovery times as a function of electric field for various excitation fluences at 12 K. The dashed curves show fits used to determine the diffusivity and mobility using Eq. (\ref{eq:tau}). (d) Schematics showing the spreading and drift of the laser pulse-injected charged carriers (top) and excitons (bottom) under electric field. (f) The  $(D/\mu)/(k_{\mathrm{B}}T/q)$ ratio at various temperatures and fluences, expected to be at unity for free charge carriers. }
     \label{fig:Ultrafast}
 \end{figure}

To complement the spatial photocurrent mapping, we performed time-resolved measurements using ultrafast transient photovoltage (TPV) to investigate the transverse-field effect on photocarrier transport with high temporal resolution. TPV, complementary to all-optical pump-probe techniques, can reveal key information on photoexcitation relaxation dynamics and transport at picosecond timescales~\cite{zeng2023ultrafast, sun2012ultrafast, massicotte2016picosecond, yagodkin2023probing, becerra2025superdiffusion, wang2026ultrafast}. In this technique, time-delayed pump and probe pulses are both focused to a 1.9 $\mu$m-diameter spot at the same location [Figure \ref{fig:Ultrafast}(a)]. At zero delay, the pump- and probe-induced carriers overlap and the local carrier density is highest, suppressing the photovoltage due to nonlinear carrier recombination.  As the delay increases, the pump-induced carriers spread out of the excitation spot and dilute the local carrier density there; the subsequently injected probe carriers then experience a lower background density, and the photovoltage recovers [Figure \ref{fig:Ultrafast}(d)]. The recovery time therefore measures the transit time for carriers to leave the excitation spot. This transit time is set by diffusion alone for charge-neutral excitons, but is shortened by drift once a transverse field acts on free carriers. The field dependence of the recovery time therefore directly distinguishes the two cases. Using this technique, we previously demonstrated anomalously high photocarrier diffusivity in Sb-doped Bi$_2$Se$_3$ nanoribbons.\cite{becerra2025superdiffusion} Here we apply it to confirm the charge-neutral nature of the photocarriers and to track their dissociation as fluence and temperature are increased.

A 200-fs, 80-MHz pulsed laser was split into time-delayed pump and probe beams, with the probe-induced photovoltage measured by lock-in detection (see SI for details). With both pulses focused at the same location at the center of the transverse-field region, we measured the open-circuit photovoltage as a function of delay time, for low- and high-fluence conditions shown in Figure \ref{fig:Ultrafast}(b) and (e), respectively. At zero delay, the photovoltage is reduced by 7\% under a pump fluence of 14 $\mu$J/cm$^2$ and by 60\% under 93 $\mu$J/cm$^2$, recovering after a characteristic delay time. At zero transverse field, the recovery time, extracted by an exponential fit on the negative delay side where the pump precedes the probe, increases from 36.5 ps to 260 ps as the fluence increases from 14 to 93 $\mu$J/cm$^2$. The slower recovery at higher fluence is consistent with our previous work\cite{becerra2025superdiffusion} and indicates a strongly reduced diffusivity at high carrier density. Applying a transverse field markedly accelerates the recovery at high fluence: the characteristic time drops from 260 ps to 80 ps at 34 mV/$\mu$m. In striking contrast, at low fluence the recovery time is essentially unchanged as the transverse field is increased. 

To extract the exciton diffusivity and mobility on ultrafast timescales, we analyze the recovery of the TPV signal as the photoexcited population moves out of the probed region by diffusion and, in the presence of a transverse field, drift. We approximate the recovery rate as the sum of the two escape channels, 
\begin{equation}
    1/\tau = 1/\tau_\mathrm{diff} + 1/\tau_\mathrm{drift} = 4D/r^{2} + \mu E_{\mathrm{t}}/r,
    \label{eq:tau}
\end{equation}
where $r$ is the laser spot radius. The zero-field intercept of $1/\tau$ versus $E_{\mathrm{t}}$ yields $D$ while the slope yields $\mu$ [Figure~\ref{fig:Ultrafast}(c)]. At 12~K, the recovery rate is nearly independent of $E_{\mathrm{t}}$ for the two lowest fluences: the fitted slopes are consistent with zero, indicating that the photoexcited population does not measurably drift in the applied field. At 23.2~$\mu$J/cm$^{2}$, $1/\tau$ begins to rise for $E_{\mathrm{t}} > 20$~mV/$\mu$m, accompanied by a weakening of the TPV signal; we tentatively attribute this to the onset of field-induced exciton dissociation, consistent with the threshold behavior observed in the steady-state photocurrent (Figure~\ref{fig:Eth-dep}). Fits are therefore restricted to the sub-threshold regime. The complete data sets and analysis are shown in Figures S8-S14 in SI.

Figure~\ref{fig:Ultrafast}(f) shows the resulting ratio $(D/\mu)/(k_{\mathrm{B}}T/q)$, which equals unity for a nondegenerate population of free carriers obeying the Einstein relation. At elevated temperatures (80--150~K) the ratio is of order unity, while at 12~K and the lowest fluence it exceeds unity by more than an order of magnitude. Because the field dependence there is nearly flat, the extracted $\mu$ carries a large uncertainty, and the ratio is determined only to within a large confidence interval, $(D/\mu)/(k_{\mathrm{B}}T/q) = 48 \pm 35$; even at the lower edge of this interval, the Einstein relation is violated by more than an order of magnitude. The extraction carries an overall geometric prefactor that is uncertain at the factor-of-2 level (e.g., whether the relevant drift length is the spot radius or diameter), which shifts all points multiplicatively and likely accounts for the modest deviations from unity at 150~K. This common-mode systematic cannot, however, produce the order-of-magnitude, temperature- and fluence-dependent enhancement observed at low temperature.

Carrier degeneracy, which enhances $D/\mu$ above $k_{\mathrm{B}}T/q$, cannot account for this deviation: degeneracy grows with carrier density and would therefore predict the strongest violation at the highest fluence, opposite to the observed trend. Instead, the violation is maximal precisely where excitonic binding is expected to be most robust---at the lowest temperature and lowest photocarrier density---and relaxes toward unity as either temperature or fluence increases, mirroring the reduction of $E_{\mathrm{th}}$ with temperature and fluence in Figure~\ref{fig:Eth-dep}. The agreement between two independent probes, steady-state SPCM and ultrafast TPV, in both the direction and the conditions of these trends supports the conclusion that the long-range transport is carried by charge-neutral excitons. 

Finally, we performed ultrafast transient photovoltage microscopy (TPVM) under transverse fields, fixing the pump at the center of the transverse field region and raster-scanning the probe beam around it.\cite{becerra2025superdiffusion} At high fluence, the photovoltage-suppression peak drifts under bias in the direction that a positive charge would follow; reversing direction with field polarity, at a rate corresponding to a mobility of $\sim$460 cm$^2$/(Vs), as expected for free carriers. At low fluence, no clear field-induced drift is resolved, consistent with charge-neutral excitons, although the weaker signal at low fluence precludes a definitive conclusion (Figure S15 in SI).

In summary, we have used a transverse electric field as a charge-sensitive probe of the long-range photoresponse in bulk-insulating Sb-doped Bi$_2$Se$_3$ nanoribbons, combining scanning photocurrent microscopy with ultrafast transient photovoltage. SPCM reveals that suppressing the photocurrent requires a transverse field of $\sim$33 mV/$\mu$m, nearly 50 times larger than the free-carrier benchmark set by the competition between transverse drift and diffusion, which depends only on device geometry and temperature. The suppression is a bias-polarity-symmetric shift of the photocurrent, uniform in laser position and mirrored by an equal enhancement on the opposite side of the contacts, reproduced by a multiple-pass exciton diffusion model in which the biased region field-ionizes excitons that repeatedly traverse the channel. Increasing temperature or fluence reduces the threshold field for the current shift, in agreement with the exciton dissociation picture.  Independently, TPV measurements show that at 12 K and low fluence the photocarrier diffusivity exceeds the value implied by the measured drift mobility through the Einstein relation by more than one order of magnitude even at the lower end of the fitting uncertainty; this violation disappears at elevated temperature or fluence, where the enhanced thermal energy and carrier screening dissociate the excitons into free carriers. Both probes converge on the same conclusion: the dominant long-range photoresponse in this system is not carried by free charges but by a charge-neutral, correlated electron-hole state, consistent with an exciton condensate. More broadly, the transverse-field threshold and the diffusivity-mobility decoupling constitute a transport-based diagnostic of charge neutrality that complements Hall- and counterflow-based signatures and can be readily applied to other candidate excitonic platforms.

\begin{acknowledgement}

This work was supported by the U.S. National Science Foundation Grants No. DMR-2404957 and No. DMR-2209884. Part of this study was performed at the UC Davis Center for Nano and Micro Manufacturing (CNM2). We acknowledge P. Klavins for his assistance in maintaining the helium recovery and liquefaction system, and Jerry Li, Tyler Valenti, Jay Huang for assisting with the experiments and simulation.

\end{acknowledgement}

\begin{suppinfo}

Experimental details, temperature-dependent photocurrent profiles under transverse field, linear plots of photovoltage versus transverse field, device characteristics, exclusion of potential artifacts, free-carrier modeling, multiple-pass model sensitivity to exciton diffusion length, complete transient photovoltage data and analysis, and transient photovoltage microscopy results. 

\end{suppinfo}

\bibliography{field-effect-BiSe}

\end{document}